\documentclass{PoS}
\usepackage{amsmath}

\title{A possible new phase in non-perturbatively gauge-fixed Yang-Mills theory}

\ShortTitle{A possible new phase in non-perturbatively gauge-fixed Yang-Mills theory}

\author{\speaker{Maarten Golterman}%
         \\
         Institut de F\'\i sica d'Altes Energies (IFAE), Universitat Aut\`onoma de Barcelona,\\ E-08193 Bellaterra (Barcelona), Spain \\and\\
        Department of Physics and Astronomy, San Francisco State University, \\San Francisco, CA 94132, USA\\
}

\author{Yigal Shamir\\
        School of Physics and Astronomy,
Raymond and Beverly Sackler Faculty of Exact Sciences,\\
Tel-Aviv University, Ramat~Aviv,~69978~ISRAEL\\
}

\abstract{The standard expectation is that gauge fixing cannot alter the physics in the physical sector of a Yang-Mills theory. Here we argue that this may not always be true: in an $SU(2)$ Yang-Mills theory in which the $SU(2)/U(1)$ coset is non-perturbatively gauge fixed, we find that a new phase, with spontaneous symmetry breaking and a Higgs-like mechanism, appears to be a possibility.}

\FullConference{31st International Symposium on Lattice Field Theory LATTICE 2013\\
                 July 29 -- August 3, 2013\\
                 Mainz, Germany}

\begin{document}

\def\floatcaption#1#2{ \caption{#2 \label{#1}} }
\def\beq{\begin{equation}}
\def\eeq{\end{equation}}
\def\bc{{\overline{c}}}
\def\bC{{\overline{C}}}
\def\tr{{\mbox{tr}}}
\def\tg{{\tilde{g}}}
\def\tA{{\tilde{A}}}

\section{Introduction}
A covariantly gauge-fixed Yang--Mills (YM) theory is defined by two
parameters, the transverse gauge coupling $g$, and the gauge-fixing
parameter $\xi$, or, equivalently, the longitudinal gauge coupling
$\tilde g$ defined by $\tilde g^2=\xi g^2$.   In perturbation theory,
both $g$ and $\tilde g$ are asymptotically free.   Of course, the
beta function for $g$ is independent of $\tilde g$, and usually one is
not concerned with the dynamics associated with the coupling $\tilde g$.
BRST symmetry guarantees that one can consistently
define a projection onto gauge-invariant correlation functions, and that
the physics defined by these correlation functions is independent of
the coupling $\tilde g$.
Nevertheless, the fact that $\tilde g$ is also asymptotically free
suggests that dynamics in the longtitudinal sector may be non-trivial.
Here, the question we will ask is whether these heuristic
observations can be put on a solid footing, and, if so, whether
the standard lore, that the physics of the transverse degrees of
freedom is always independent of what happens in the longitudinal
sector, is indeed inescapable.   We will argue that it is not \cite{GS2012}.

The first task is to formulate the gauge-fixed theory non-perturbatively.
We start from a YM theory on the lattice, where no gauge fixing
is required.   One may then try to gauge-fix this theory
rigorously following the textbook procedure, imposing BRST symmetry
of the gauge-fixed theory on the lattice \cite{HN1986}.   However, it was
found that this does not work:  the gauge-fixed partition function (as well
as the unnormalized expectation values of gauge-invariant operators)
vanishes identically, as can be proven using BRST symmetry
\cite{HN1987}.   The physical explanation is that the presence of
Gribov copies in the gauge-fixed theory allows the integrand of the
path integral to become negative, with BRST symmetry enforcing an
exact cancelation.   In fact, this already happens if one integrates
only over the gauge degrees of freedom and the ghosts, while keeping
the transverse gauge field fixed, {\it i.e.}, it happens on each orbit
in the space of gauge-field configurations.

The no-go theorem of Ref.~\cite{HN1987} can be avoided if we only partially
fix the gauge, leaving a subgroup $H$ of the gauge group $G$ containing minimally the Cartan subgroup unfixed \cite{GS2012,MS,GS2004,FF2013}.   We will
review this observation, for the case $G=SU(2)$ and $H=U(1)$, in Sec.~2
below.   We will see that the restriction of this ``equivariantly'' gauge-fixed
theory to the trivial orbit is an example of a topological field theory (TFT).
The interesting question then arises whether spontaneous symmetry breaking can take place in a TFT.   In Sec. 3, we show that in a toy model, this is indeed the case.   We then report on the potential consequences of this observation for the gauge-fixed $SU(2)$ YM theory in Sec. 4, arguing that, contrary to standard lore, there is no
logical argument excluding a phase in which the longitudinal dynamics does
in fact affect the physics of the theory.   In the mean-field approximation, we find evidence that such a phase may indeed exist.
Assuming that such a phase exists, we discuss future explorations of this novel
scenario in Sec.~5.

\begin{boldmath}
\section{Equivariant BRST symmetry and the $SU(2)/U(1)$ coset theory}
\end{boldmath}
Standard gauge fixing works by inserting into the unfixed $SU(2)$ YM
theory, with partition function
\beq
\label{Zun}
Z=\int[dU]\,\mbox{exp}[-S_{\rm inv}(U)] ,
\eeq
the constant
\beq
\label{Zgfstandard}
\mbox{constant}=Z_{\rm gf}(U,\xi)=\int[d\phi][dc][d\bc]\,\mbox{exp}[-
S_{\rm gf}(U^\phi,c,\bc)]\ .
\eeq
Here $\xi$ is the gauge-fixing parameter, $\phi_x\in SU(2)$,
and the gauge transform of $U_{x, \mu}$ is
$U^\phi_{x, \mu} = \phi_x U_{x, \mu} \phi^\dagger_{x+\hat\mu}$.
Using BRST invariance of the gauge-fixing action $S_{\rm gf}$
and nilpotency of the BRST transformations one can prove
that indeed $Z_{\rm gf}(U,\xi)=\mbox{constant}$, {\it i.e.}, independent of $U$ and
$\xi$.   Unfortunately, this constant takes the one
value that is not allowed: it vanishes \cite{HN1987}.

In equivariant gauge fixing, only the coset $SU(2)/U(1)$ is gauge-fixed.\footnote{For $SU(2)$ this is the only nontrivial coset; for larger groups
there are more possibilities \cite{GS2012,GS2004}.}   Taking the $U(1)$
subgroup to be generated by the third Pauli matrix $\tau_3$, we introduce
ghosts $C$ (and anti-ghosts $\bC$) that are coset valued, {\it i.e.},
\beq
\label{cosetgh}
C=C_1\tau_1/2+C_2\tau_2/2\ ,\qquad\bC=\bC_1\tau_1/2+\bC_2\tau_2/2\ ,
\eeq
and we impose the new BRST transformation $sC=0$
instead of $sC=-iC^2$, which is proportional to $\tau_3$, and would thus
take the transformed $C$ field out of the coset.   It is possible to extend the
definition of this new BRST transformation, to which we will refer as
``equivariant'' BRST (eBRST) to the other fields such that for all fields
$s^2(\mbox{field})=\delta_{U(1)}(\mbox{field})$,
with $\delta_{U(1)}$ a $U(1)$ gauge transformation (with parameter
$\propto C^2$),
so that $s$ is nilpotent on any operator invariant under the unfixed group
$U(1)$.   A consequence is that, in order to maintain invariance under
eBRST transformations, the gauge-fixing action has to be of the form
\beq
\label{Sgf}
S_{\rm gf}(U^\phi,C,\bC)=\frac{1}{\xi g^2}\,\tr\left(F(U^\phi)\right)^2+2\tr(\bC M(U^\phi)C)-2\xi g^2\,\tr\left(C^2\bC^2\right)\ .
\eeq
Here $F(U)$ represents the choice of gauge, and $M(U)$ is the corresponding Faddeev--Popov operator.   The first two terms are also present in the standard case, but the four-ghost term is new, and essential in order to
maintain invariance under eBRST.

Insisting on Lorentz invariance, local $U(1)$ invariance,
and (power-counting) renormalizability, the choice of $F(U)$ is essentially unique.   In continuum language, we choose
\beq
\label{gfchoice}
F\sim D_\mu(A)W_\mu\ ,\qquad\ V_\mu=\frac{1}{2}\left(W^1_\mu\tau_1
+W^2_\mu\tau_2+A_\mu\tau_3\right) ,
\eeq
with $U_\mu=\mbox{exp}(iV_\mu)$, and where $D_\mu(A)$ is the
covariant derivative with respect to the $U(1)$ subgroup.   For all details,
we refer to Refs.~\cite{GS2012,GS2004}.

The key observation is that now the constant $Z_{\rm gf}(U,\xi g^2)$
does not depend on $U$ or $\tg^2=\xi g^2$ but also does not vanish.
For any $U$, the path integral $Z_{\rm gf}(U,\xi g^2)\ne 0$ defines a TFT.
This leads to an {\it invariance} theorem \cite{MS,GS2004},
\beq
\label{invth}
\langle{\cal O}(U)\rangle_{\rm unfixed}=\langle{\cal O}(U)\rangle_{\rm eBRST}\ ,
\eeq
for any gauge-invariant operator $\cal O$.
The eBRST gauge-fixed theory is rigorously the same
as the unfixed theory when restricted to gauge-invariant correlation functions.

In possession of a non-perturbatively well-defined gauge-fixed YM theory,
we can now ask the question whether the gauge-fixing sector can have
an effect on the physics, in contrast to the standard lore.
We begin by considering the TFT on the trivial orbit, {\it i.e.}, the theory defined by
\begin{eqnarray}
\label{reduced}
Z_{\rm gf}(1,\tg^2)&=&\int[d\phi][dC][d\bC]\,\mbox{exp}[-S_{\rm gf}(1^\phi,C,\bC)]\ ,\\
S_{\rm gf}&=&\frac{1}{\tg^2}\,\tr\left(F(1^\phi)\right)^2+2\,\tr(\bC M(1^\phi)C)-2\tg^2\,\tr\left(C^2\bC^2\right)\ ,\nonumber
\end{eqnarray}
with $1^\phi=\phi_x\phi^\dagger_{x+\mu}$ (on the lattice).  This model,
the ``reduced model,'' is in itself a strongly
interacting theory with an asymptotically free coupling $\tg$ \cite{GS2005}.
It is still invariant under an eBRST symmetry (in which $\phi$
transforms as $s\phi=-iC\phi$, and $U=1$ does not), and moreover it is invariant under
\beq
\label{symred}
\phi_x\to h_x\phi_x g^\dagger\ ,
\eeq
with $h_x\in U(1)_L$ a local symmetry projecting $\phi_x$ onto the coset $SU(2)/U(1)$,
and $g\in SU(2)_R$ a global symmetry (reminiscent of a custodial symmetry).   Again, for details we refer to Ref.~\cite{GS2012}.

Forgetting for the moment that our reduced model is a TFT, one might
ask whether, because of the strong dynamics, SSB could take place.   A possible order parameter is $\langle\phi^\dagger\tau^3\phi\rangle$, which is invariant under the
gauge symmetry $U(1)_L$, but for which a non-zero value would signal SSB along the pattern
$SU(2)_R\to U(1)_R$.   This would correspond to a non-trivial minimum
of the effective potential
\beq
\label{Veff}
\mbox{exp}[-V_{\rm eff}(\tA)]=\int[d\phi][dC][d\bC]\,
\delta\left(\tA-\frac{1}{V}\sum_x\phi^\dagger_x\tau_3\phi_x\right)\,
\mbox{exp}[-S_{\rm gf}(1^\phi,C,\bC)]\ .
\eeq
However, remembering that $Z_{\rm gf}$ is a TFT, an apparent paradox arises: the question is whether
$V_{\rm eff}(\tA)$ can be non-trivial, given that $dZ_{\rm gf}/d\tg=0$.
In other words, can SSB ever take place in a TFT?   In order to address this
question, we first consider a toy model.

\section{A toy model}
Let us consider the simple zero-dimensional field theory (a.k.a.\ integral)
\beq
\label{toy}
Z=\frac{1}{2\sqrt{\pi}}\int_{-\infty}^\infty d\phi\int dcd\bc\,\mbox{exp}[-f^2(\phi)/4+\bc f'(\phi)c]\ ,
\eeq
with $f$ an arbitrary function going to $+\infty$ ($-\infty$) for $\phi\to+\infty$
($-\infty$).   This theory is invariant under a ``baby'' BRST symmetry
$s\phi=c$, $sc=0$, $s\bc=f(\phi)/2$.   By straight evaluation,
we have that $Z=1$.  The toy model is a ``TFT'' because $Z$ does not depend
on the choice of the function $f$.

Consider a specific example
\beq
\label{fchoice}
f(\phi)=\frac{1}{\lambda}\left(\phi^3-v^2\phi\right)\ ,\qquad\lambda,\ v\ \mbox{real}\ .
\eeq
The theory is now also invariant under a $Z_2$ symmetry $\phi\to-\phi$.
The minima of the ``classical potential'' $f^2(\phi)/4$ are at $v=0$ (unbroken $Z_2$) and $\phi=\pm v$
(SSB of $Z_2$).   However, the invariance theorem would seem to imply
that always $\langle\phi\rangle=0$, and therefore $Z_2$ can never be broken!

Let us recall how one studies SSB.   We first turn on a small, explicit
symmetry-breaking term; here we will choose $S_{\rm seed}=-\epsilon\phi$.
We then take the infinite-volume limit, and only after that do we take $\epsilon\to 0$.   The seed breaks not only $Z_2$, but also the baby BRST symmetry of the toy model, which therefore at $\epsilon\ne 0$ is no longer a TFT.   Of course, in our zero-dimensional field theory we cannot take the volume to
infinity, but we can define different ``vacua'' by considering the perturbative
expansion in $\lambda$ around each of the three saddle points $\phi=\pm v,\ 0$.
Using a subscript taking values $\pm v,0$ to indicate quantities calculated
in perturbation theory around a given saddle point,
we find (for $\epsilon\to 0$) that
\beq
\label{vev}
\langle\phi\rangle_v=v\left(1-\frac{3}{4}\frac{\lambda^2}{v^6}+\dots\right)\ .
\eeq
We also find that  $Z_{\pm v}=1$ and $Z_0=-1$ to all orders in $\lambda$,
which sum up to $Z=1$.

For $\epsilon>0$, the saddle-point approximation for $Z$, valid for small $\lambda$, prefers the minimum at $\phi=v$, thus breaking the symmetry.
In any dimension larger than zero, the contribution of only one of these saddle point would survive the thermodynamic limit,
and for $\epsilon \searrow 0$ it would be the one at $\phi=v$.
The conclusion is that, for $v^2>0$ in Eq.~(\ref{fchoice}),
indeed the toy model teaches us that SSB can take place in a TFT!
The lesson is that in order to probe SSB, a ``seed" breaking the symmetry
explicitly has to be added to the theory.   With the seed present, the
invariance theorem no longer applies, and it may or may not apply after
appropriate limits have been taken.

The baby BRST symmetry remains unbroken around each of the saddle points.
An order parameter for SSB of BRST would have the form
$sX$ for some operator $X$, but one can prove that
$\langle sX\rangle_v$ vanishes for all $X$ (apart from terms proportional to
$\epsilon$).

\section{The phase diagram in mean field}
Now let us return to the reduced model of Eq.~(\ref{reduced}).   We turn on a
``magnetic'' field $h$ in the $\tau_3$ direction, coupled to the order parameter
of Sec.~2:
\beq
\label{seed}
S_{\rm seed}=-h\,\tr\left(\tau_3\phi^\dagger\tau_3\phi\right)\ .
\eeq
This addition breaks $SU(2)_R\to U(1)_R$, and breaks eBRST
symmetry, while respecting the $U(1)_L$ unfixed gauge invariance.
Therefore, for $h\ne 0$ the invariance theorem does not apply.   Whether
it applies after taking first the infinite-volume limit and then $h\to 0$ is a
{\it dynamical} question!

It is also a non-perturbative question.   In Ref.~\cite{GS2012} we studied
this question by first integrating out the ghosts in a $1/\tg^2$ expansion,
and then applying a mean-field approximation to the resulting effective
action $S_{\rm eff}(\phi)$.   In this approximation, we find that
\begin{enumerate}
\item Starting from the strong-coupling limit $\tg=\infty$
there is a first-order phase transition at $\tg=\tg_c$, with
\beq
\label{phtr}
\langle\phi^\dagger\tau_3\phi\rangle\
\begin{cases}=0\ ,\quad \tilde{g}>\tilde{g}_c\ , \\
\ne 0\ ,\quad \tilde{g}<\tilde{g}_c\ .
\end{cases}
\eeq
\item When the gauge fields $U$ are made dynamical again
(by promoting $SU(2)_R$ to a local symmetry),
the Goldstone bosons associated with the $SU(2)_R\to U(1)_R$ symmetry breaking provide
longitudinal polarizations for the $W$ gauge fields of Eq.~(\ref{gfchoice}),
which become massive, while the ``photon'' (the $A$ field) stays massless.
\item We do not know the fate of eBRST symmetry.   But we observe that
$\phi^\dagger\tau_3\phi$ is not the eBRST variation of anything, so a
vacuum expectation value for this operator does not imply SSB of eBRST.
We also note that the $W$'s picking up a mass is not inconsistent with
unbroken eBRST symmetry if also the ghosts pick up an equal mass.
The mass term
\beq
\label{mass}
S_{\rm mass} =m_W^2\int d^4x\,\tr\left(W_\mu^2+2\bC C\right)
\eeq
is eBRST invariant.
\end{enumerate}
%%%%%%%%%%%%%%%%%%%
%%% diagrams
\begin{figure}
\includegraphics*[width=7.0cm]{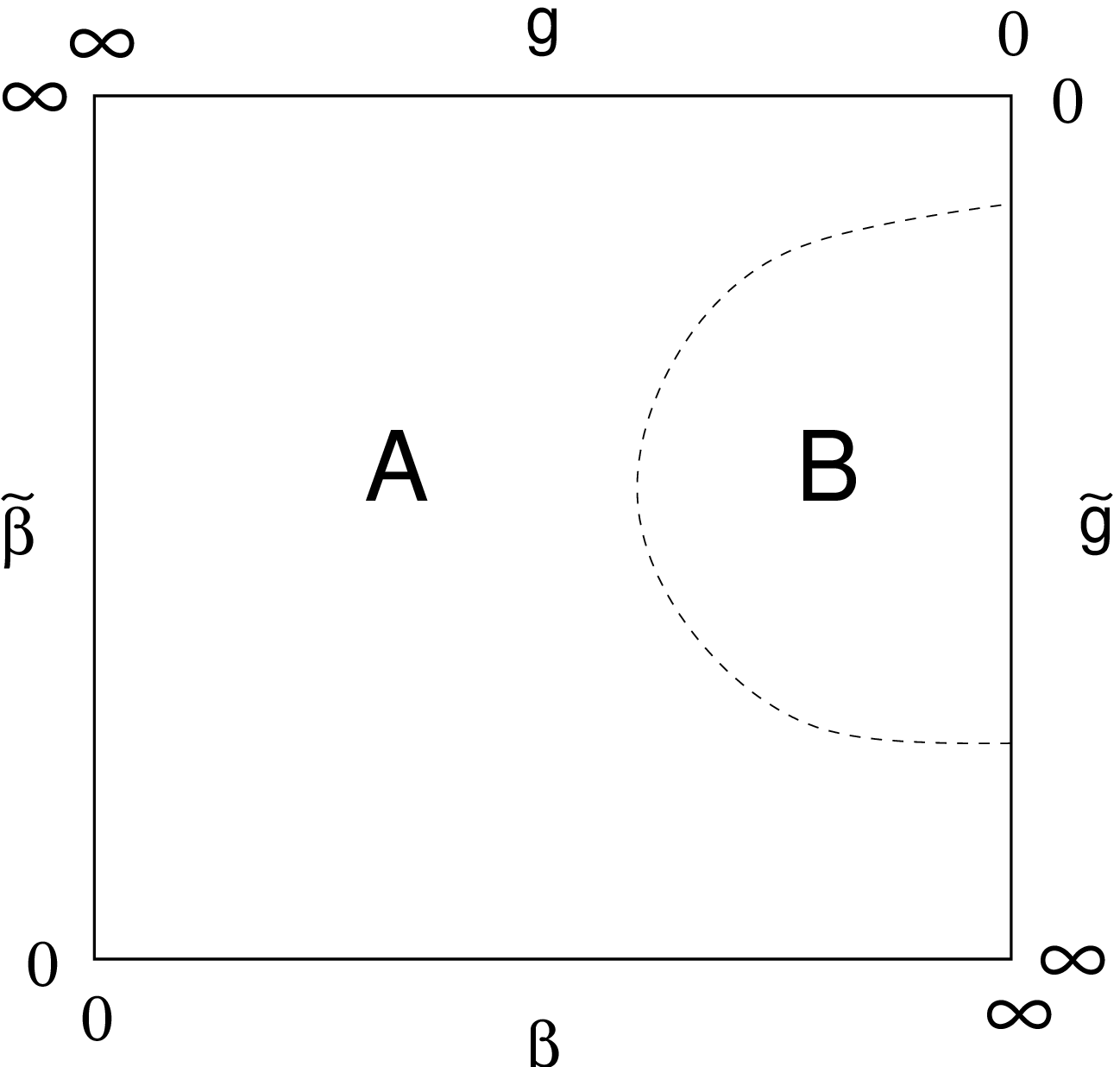}
\hspace{1cm}
\includegraphics*[width=7.0cm]{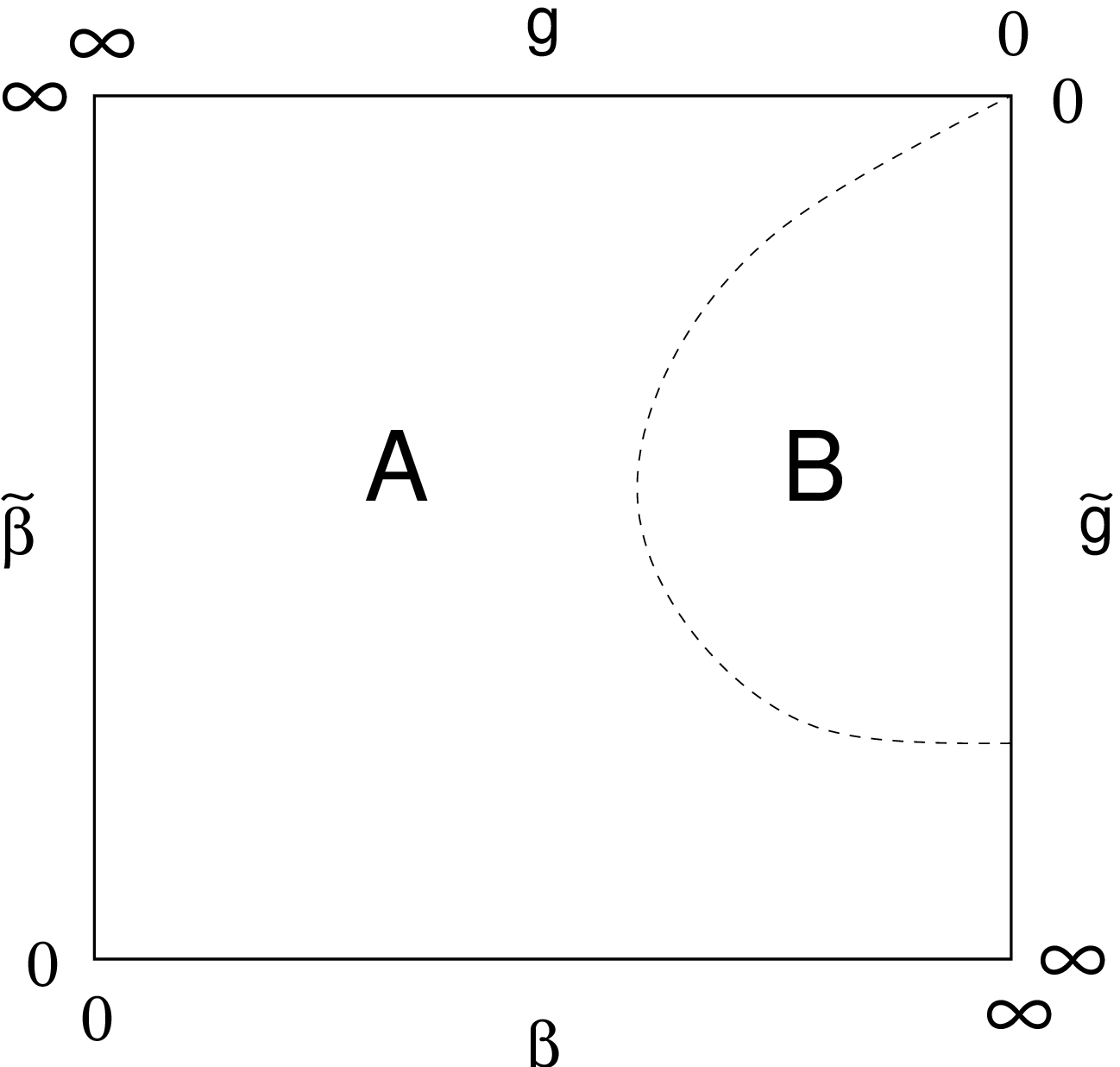}
\vspace*{-2ex}
\begin{quotation}
\floatcaption{phasediag}{%
Two scenarios for the phase diagram.  The confining phase A has a mass gap,
while the Coulomb (or Higgs) phase B has a massless photon.
Left panel: the Coulomb phase ends at some non-zero $\tg$ for $g\to 0$.
Right panel: the Coulomb phase extends to the critical point at $g=\tg=0$.}
\end{quotation}
\vspace*{-3ex}
\end{figure}
%%%%%%%%%%%%%%%%%%%
The consequences of our results in the reduced model for the full phase diagram are as follows.   The reduced-model transition is shown in
Fig.~\ref{phasediag} on the line $g=0$; it corresponds to the value
nearest to $\tg=\infty$.   This transition extends into the diagram, separating a
confining phase A, from a Higgs-like phase B.   In phase A there is a mass
gap, as in the unfixed $SU(2)$ YM theory, while phase B does not have a
mass gap, because the photon remains massless.   There is thus a
clear phase separation also in the full theory.   Semi-rigorous arguments
on the other edges of the phase diagram suggest that the phase transition
line has to end up on the $g=0$ edge, either ending at $\tg>0$ (left panel)
or ending at $\tg=0$ (right panel) \cite{GS2012}.

If the real phase diagram would look like the left panel, the whole new
phase B would be a lattice artifact, since the continuum limit is at
$g=0$, $\tg=0$.  If, however, it would look like the right panel, this would
imply the existence of a completely new continuum theory, with Higgs-like
properties, but no fundamental scalars, and with only asymptotically free
couplings.   In order to reach this continuum limit, one should take the limit from within phase B, whereas the continuum limit from within phase A would
lead to the usual confining phase of $SU(2)$ YM theory.

We do not know which of these two scenarios may apply, if any.   However,
our study of the one-loop RG equations of Ref.~\cite{GS2005} suggests
that there are two different regimes.   Each of the asymptotically free
couplings $g$ and $\tg$ has a scale associated to it, $\Lambda$, respectively
$\tilde\Lambda$.   The RG equations lead to two types of solutions:
either $\Lambda\approx\tilde\Lambda$, or $\Lambda\ll\tilde\Lambda$.
We conjecture that $\Lambda\approx\tilde\Lambda$, where the transverse
degrees of freedom dominate, is to be identified
with the usual confinement phase A.  The other situation,
$\Lambda\ll\tilde\Lambda$, corresponds
to taking the continuum limit
well inside phase B, where the bare $\tg$ is much larger than the bare $g$.
In this case, the non-perturbative physics associated with $\tilde\Lambda$ would dominate, while $g$ is still
perturbative.   Combined with our mean-field solution, this suggests the
possibility that in that case a $W$-boson mass $m_W\sim g\tilde\Lambda$
gets generated.

\section{Outlook}
The scenario we presented raises a number of important questions:
\begin{enumerate}
\item Does the new phase B exist?   A numerical study answering this may well be possible.
\item If it exists, does it extend to $g=\tg=0$?  Here analytic studies will be
necessary, and we are hopeful that a combination of small-volume and
large-$N$ techniques (applied to the case $SU(N)/[SU(N-1)\times U(1)]$)
may help.
\item Is the continuum limit in phase B unitary, if phase B exists and extends to the gaussian fixed point?  Is this the same as asking
whether eBRST stays unbroken?
\item What distinguishes the two phases microscopically in the full theory?
In the full theory, the seed should break eBRST in order to avoid the invariance theorem, and this may bias the sum over
Gribov copies in the equivariantly gauge-fixed theory.
\end{enumerate}

What we did show so far is that a scenario in which spontaneous symmetry breaking takes place in a topological field theory is a logically consistent
scenario.   Despite the fact that the partition function of a topological field theory does not depend on its couplings, and would thus not appear to have
any phase structure, we showed that this conclusion is not necessarily valid.

What we do not know is whether in equivariantly gauge-fixed $SU(2)$ YM
indeed a new phase, originating from the spontaneous symmetry breaking
in the trivial-orbit theory (the reduced model), exists, even though mean
field suggest it may.   But if the reduced model does undergo spontaneous
symmetry breaking, the existence of a phase with a Higgs-like character
(massive $W$'s and a massless photon) appears inescapable, since
the Goldstone bosons will combine with the $W$ fields to make them massive.\\

{\bf Acknowledgements}
YS thanks the Department of Physics and Astronomy of San Francisco
State University for hospitality.  This work was
supported in part by the US Department of Energy, and by the Spanish Ministerio de Educaci\'on, Cultura y Deporte, under program SAB2011-0074 (MG), as well as
by the Israel Science Foundation under grant no.~423/09 (YS).
We also thank the Galileo Galilei Institute for Theoretical Physics for
hospitality, and the INFN for partial support.

\end{document}